# Investigation of Countercurrent Flow Profile and Liquid Holdup in Random Packed Column

Yucheng Fu[1,*], Jie Bao[2], Rajesh Singh[2], Chao Wang[1], Zhijie Xu[1]

[1]Physical and Computational Sciences Directorate,

Pacific Northwest National Laboratory, Richland, WA 99352, USA

[2]Energy and Environment Directorate,

Pacific Northwest National Laboratory, Richland, WA 99352, USA

**Abstract**

Liquid holdup and mass transfer area are critical parameters for packed column design and $CO_2$ capture efficiency prediction. In this paper, a framework was established for modeling the liquid–gas countercurrent flow hydrodynamics in a random packed column with pall rings. Besides the column-averaged information, the radial pall ring distribution, velocity, and liquid holdup profiles are obtained to study the entrance effect and the wall influence in the packed column. With local CFD data, the packing specific area $a_p$ and liquid velocity $u_L$ range are significantly expanded for liquid holdup correlation development with even only one size pall rings. The proposed liquid holdup correlation $h_L \sim u_L^{0.44}$ indicates the random packed column falls in a viscous to turbulent transition regime with a Reynolds Number range of [6.7–40.2]. The derived liquid holdup correlation is in good agreement with existing correlations developed using the column-averaged experimental data.

## 1 Introduction

Carbon dioxide ($CO_2$) generation from fossil fuel combustion is considered to be a significant contributor to climate change [1]. Capturing $CO_2$ from large point sources, transporting it to a storage site, and geologically depositing it is the most effective approach to reduce $CO_2$ emissions [2]. In $CO_2$ capture, post-combustion capture is the most popular configuration because existing fossil fuel power plants can be retrofitted to include $CO_2$ capture technology[3].

[*] Corresponding Author: Yucheng Fu (email: yucheng.fu@pnnl.gov)

In the post-combustion process, the packed column is commonly used for $CO_2$ capture by providing an enhanced contact area for the gas and liquid phases. The study of mass transfer area and liquid holdup behavior is critical to improve the packed column performance and the $CO_2$ capture efficiency. Numerous experimental data and various correlations have been developed for random packed columns[4]–[13]. Column-averaged flow measurement and prediction is generally straightforward and efficient to determine using existing experimental methods. For example, liquid holdup $h_L$ can be measured by shutting off the liquid inlet and outlet simultaneously and weighing the liquid amount in the column. The superficial liquid velocity $u_L$ and gas velocity $u_G$ can be measured using a flow meter with a known column diameter. For the radial $h_L$ profile measurement, the equally distributed bottom annular sampling container [6], electrical resistance tomography system [14], or gamma-ray/X-ray system can be used for measurement [15],[16]. The mass transfer area is not directly accessible in experiments. Instead, the effective mass transfer area is backed out based on the measured $CO_2$ concentration at the inlet and outlet with known mass transfer coefficients [17]–[19]. These methods generally have relatively low spatial resolution and need to be customized for different packed column systems.

Compared to experimental methods, the 3-dimensional (3D) multiphase CFD simulation can provide high-resolution data for all the aforementioned quantities in the complex random packed columns. Extensive CFD studies have been carried out for studying countercurrent flow hydrodynamics in packed columns [20]–[26]. In the literature, most of the existing CFD models focused on the packed columns with less than 100 packing elements. The common packing elements used in the column are spheres, cylinders, and Raschig rings [27]. For complex geometry such as pall ring or multi-hole cylinders, the available studies are mostly focused on packing ring distribution with limited hydrodynamics data [28],[29].

This study seeks to develop a general framework for modeling 3D countercurrent flow in a random packed column using pall rings. The high-resolution simulation data are used to study pall ring distribution and local countercurrent flow characteristics. The impacts of ring distribution inhomogeneity on solvent distribution and velocity are investigated. Since the laboratory-scale 3D multiphase CFD simulation is computationally expensive, only limited runs can be carried out. This creates a challenging task for hydrodynamic correlation development in the packed column. The reason is that these correlations normally require a sufficient amount of available data, especially when the objective parameter, such as liquid holdup or gas–liquid interfacial area, depends on more than one parameter. This paper attempts to address this issue by breaking down all laboratory-scale CFD high-resolution simulations into multiple subdomains with local flow data. The acquired countercurrent flow radial profiles increase the number of training datasets for building realistic and reliable correlations. The developed liquid holdup correlation using this method will be validated with existing correlations and column-averaged liquid holdup results. In Section 2, the CFD method and models applied in this study are briefly introduced. The geometry and boundary conditions of the CFD model are introduced in Section 3. The acquired results for the pall ring packed column are shown and discussed in Section 4.

## 2  Methodology

The STAR-CCM+, a commercial CFD software [30], was used for the 3D multiphase flow investigation in a representative model of a random packed column. The numerical model involves solving the mass and momentum conservation equations. Multiphase fluid separation is modeled by the volume of fluid (VOF) method. The mass conservation equation has the form of:

$$\frac{\partial \rho}{\partial t} + \nabla \cdot (\rho \mathbf{u}) = 0, \tag{1}$$

and the momentum equation has the form of:

$$\frac{\partial \rho \mathbf{u}}{\partial t} + \nabla \cdot (\rho \mathbf{u}\mathbf{u}) = -\nabla p + \mu \nabla^2 \mathbf{u} + \rho \mathbf{g} + \mathbf{F}_\sigma, \tag{2}$$

where $\rho$ is density, $\mu$ is viscosity, $p$ is pressure, and $\mathbf{F}_\sigma$ is the surface tension force at the gas–liquid interface. Density and viscosity are averaged by the liquid and gas phase fraction as:

$$\rho = \rho_L \alpha + \rho_G (1-\alpha),$$
$$\mu = \mu_L \alpha + \mu_G (1-\alpha), \tag{3}$$

where $\alpha$ is the void fraction of the liquid phase. The interfacial surface tension force $\mathbf{F}_\sigma$ is computed as

$$\mathbf{F}_\sigma = \sigma \kappa \nabla \alpha, \quad \kappa = -\nabla \cdot \frac{\nabla \alpha}{|\nabla \alpha|}, \tag{4}$$

where $\sigma$ is the surface tension coefficient and $\kappa$ is the local surface mean curvature. On the wall region, the $\frac{\nabla \alpha}{|\nabla \alpha|}$ is enforced as: $-n_w \cos\theta_w + t_w \sin\theta_w$. $n_w$ and $t_w$ are the unit normal and tangential vectors of the wall surface, respectively. The contact angle $\theta_w$ is where the gas–liquid interface meets the solid surface. The transport of the void fraction $\alpha$ is governed by:

$$\frac{\partial \alpha}{\partial t} + \nabla \cdot (\mathbf{u}\alpha) = 0. \tag{5}$$

## 3   Problem setup and boundary conditions

3D multiphase countercurrent flows are simulated to understand hydrodynamics in a random packed column. The column is filled with same-size pall rings. In order to generate an arbitrary random packed column, the pall rings are dumped into the cylindrical column using the discrete element method (DEM) from the commercial software STAR-CCM+[21]. The geometry of the pall ring is adapted from Chen et al. [4], with the dimensions depicted in Figure 1 (a). The diameter and height of the pall ring are 16 mm and 16 mm, respectively. The wall thickness of the pall ring is 0.5 mm. The cylindrical column has a diameter of 100 mm and a height of 200 mm. The column-to-ring diameter ratio is $D/d = 6.25$.

During the packing process, the pall ring was simplified as a solid cylinder with the same dimensions. Once settled down in the packed domain, the cylinders will be replaced by the actual pall rings using acquired cylinder locations and orientations. The computational domain can then be generated and meshed by subtracting the pall ring from the cylindrical column region. A total of 160 pall rings were randomly packed in the column in this study. The computational domain setup and final pall ring arrangement in the packed column are shown in Figure 1 (b). The distributor (marked in blue) is installed at the top of the domain and the liquid is fed into the computational domain at a given liquid load. The remaining part of the top surface is the gas outlet, as shown in yellow. The side wall of the computational domain is specified as the no-slip wall. The bottom of the flow domain is the pressure boundary and has a gradient of zero for the volume fraction of the liquid phase.

Apart from the pall ring packing, the column-top liquid distributor is another critical component in the packed column that can have a significant influence on liquid/solvent distribution. The details of the liquid distributor dimension and arrangement are shown in Figure 2. The distributor has 13 drip holes. Each drip hole has a diameter of 1 mm, and the liquid flow is directed vertically downward into the packed column. The position of the drip holes from the column center is normalized by the column radius $R = 50$ mm. The green, blue, red and yellow colors represent the normalized radial positions of those holes: $r/R = 0, 0.3, 0.42, 0.6$, respectively.

In the random packed column, column-averaged porosity $\epsilon$ and specific area $a_p$ are two important geometrical parameters that are closely related to packed column performance. The porosity $\epsilon$ is defined as the void region over the total empty column volume $V_S$. The specific area $a_p$ is defined as the ring surface area per unit volume. In this random packed column, the column-averaged porosity $\epsilon$ is calculated as 0.97 and the specific area $a_p$ is 218 $m^2/m^3$.

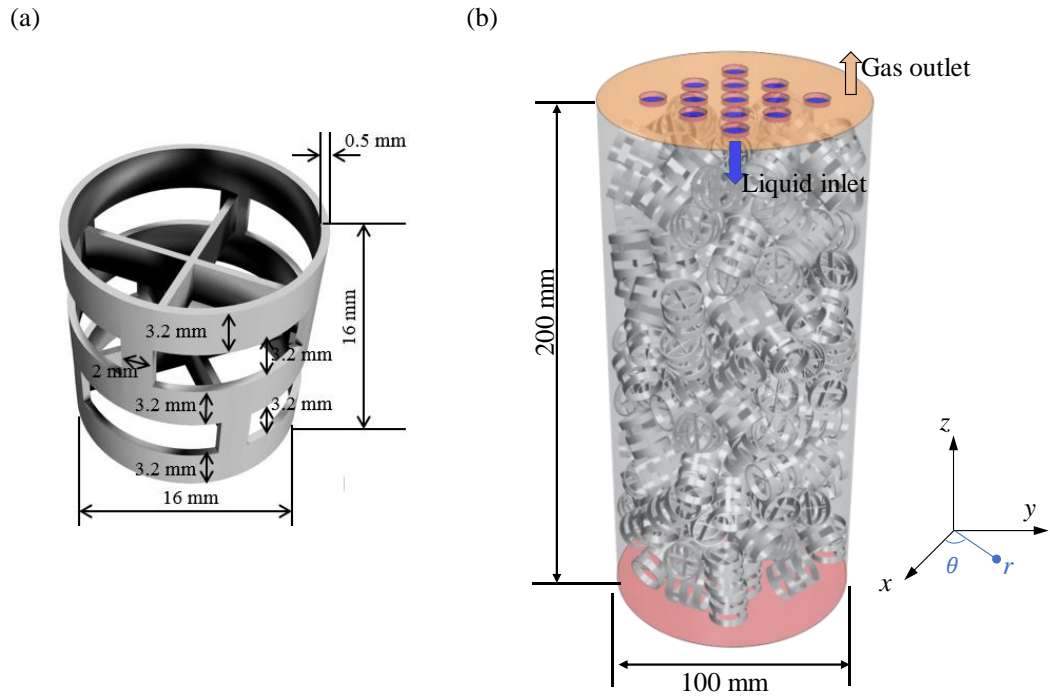

Figure 1: (a) Design and dimension details of the pall ring; (b) Schematic of the flow domain showing the arrangement of the pall rings inside the packed column with an ID 100 mm and height of 200 mm. The pall rings are dumped into the flow domain using the DEM.

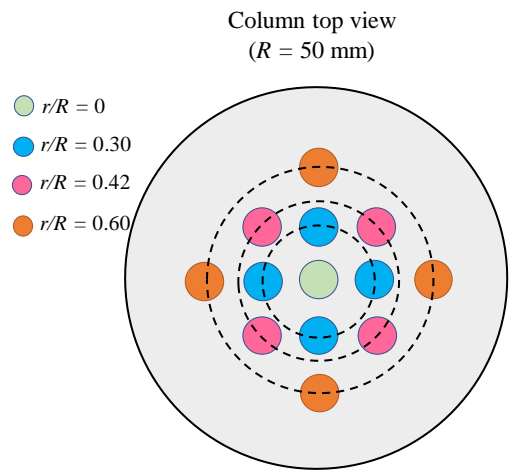

Figure 2: The arrangement of 13 liquid drip holes at the top surface of the packed column. The circle color represents four sets of liquid inlets that have the same radial distances to the column center.

The computational domain is discretized into polyhedral meshes with prism layers around all solid surfaces (pall ring surface and container wall). A snapshot of the generated mesh is shown in Figure 3. A total of 13 million mesh cells is generated using this setup. The average mesh size is 0.78 mm, with a standard deviation of 0.72 mm. The physical properties of the gas and liquid used in this study are summarized in Table 1. Flow simulations were run on the Pacific Northwest National Laboratory (PNNL) Institutional Computing (PIC) high-performance computing (HPC) cluster. Each compute node has dual Intel Haswell E5-2670 CPUs giving 24 cores per node. Each simulation ran on 96 CPU cores, and around 7 hours wall clock time of simulation can propagate a one-second solution time. All the simulations run to $T = 5s$ solution time to achieve a pseudo steady-state condition.

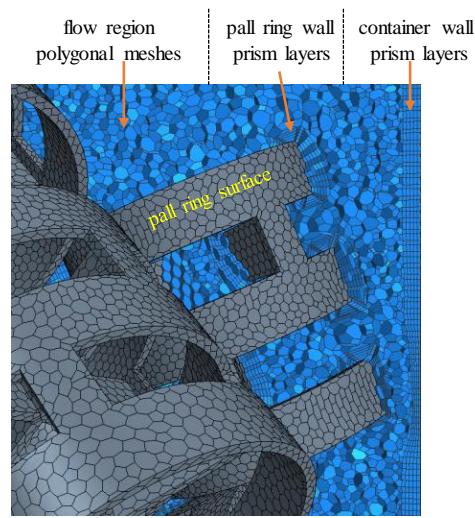

Figure 3: Snapshot of the CFD mesh generated in the computational domain. The prism layers are generated on both the pall ring surface and container wall. The polyhedral meshes are used inside the flow domain.

Table 1: Liquid and gas properties used in CFD simulations.

| Property | Notation and Unit | Value |
| --- | --- | --- |
| Liquid density | $\rho_L$, $[kg \cdot m^{-3}]$ | 1058 |
| Gas density | $\rho_G$, $[kg \cdot m^{-3}]$ | 1.0 |
| Liquid viscosity | $\mu_L$, $[Pa \cdot s]$ | 0.00246 |
| Gas viscosity | $\mu_G$, $[Pa \cdot s]$ | $1.855 \times 10^{-5}$ |
| Liquid surface tension | $\sigma$, $[N \cdot m^{-1}]$ | 0.065 |
| Contact angle | $\theta_w$ [°] | 40 |

## 4 Results and Discussion

Six liquid loads are investigated in this study, which are $q_L$ = 12.2, 24.4, 36.7, 48.9, 61.1, and 73.3 m³/m²/h. The corresponding liquid superficial velocity $u_L$ can be calculated from liquid load as:

$$u_L = q_L / 3600 \text{ [m/s]}. \tag{6}$$

These liquid load cover the Reynolds number (Re) range of [6.7–40.2], where Re is defined as:

$$\text{Re} = \frac{\rho_L u_L}{a_p \mu_L}. \tag{7}$$

The gas velocity is kept at a constant $u_G$ = 0.28 m/s.

With the above liquid and gas velocity range, no obvious liquid blockage is observed in the countercurrent flow simulation. The solvent distribution at liquid load $q_L$ = 12.2, 48.9, and 73.3 m³/m²/h is visualized in Figure 4 at $T$ = 5s. These figures are acquired at the center cross section of the packed column. At the low liquid load $q_L$ of 12.2 m³/m²/h, as shown in Figure 4 (a), the liquid occupied a relatively small portion of the flow region. Most of the liquid forms the trickle films on the pall ring surface. With the increase of the liquid load, the solvent spreads more widely and a larger portion of the ring surface area is covered by the liquid film, as shown in Figure 4 (b) and (c). Free streams are formed at high liquid load conditions.

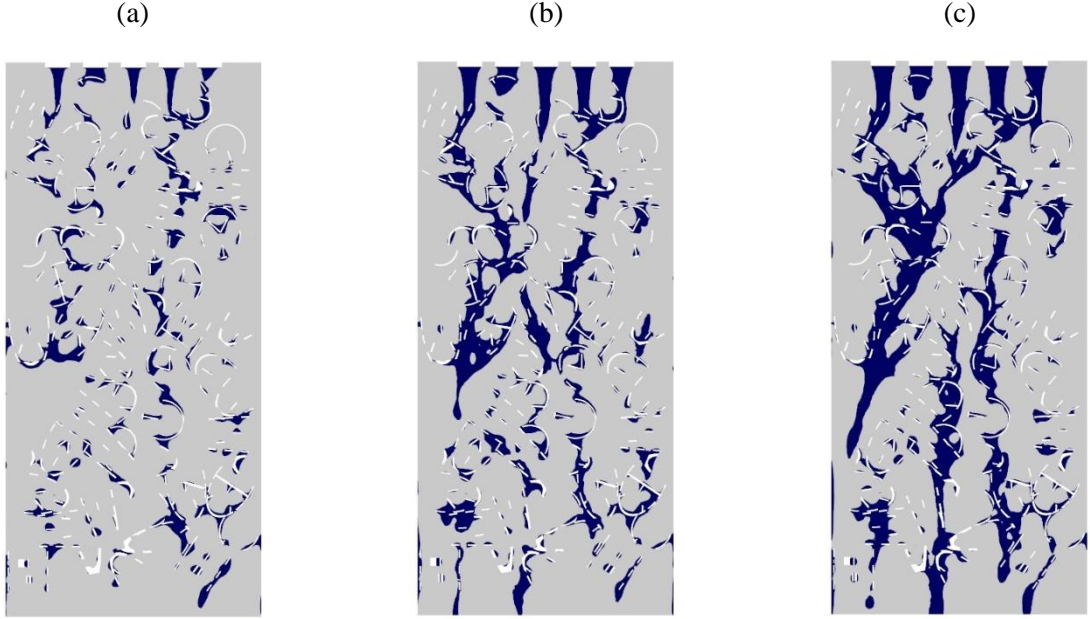

Figure 4: Solvent distribution (blue color) at the central vertical plane of the packed column at liquid load $q_L$ of (a) 12.2, (b) 48.9, and (c) 73.3 m³/m²/h at $T = 5$s.

More quantitative results are provided in the following sections. In Section 4.1, the mass transfer area acquired in CFD is compared with the experimental data for model validation. The radial pall ring distribution effect on the packed column porosity $\epsilon$ and specific area $a_p$ profiles is discussed in Section 4.2. These two geometrical parameters provide the necessary information to understand radial two-phase velocity and liquid holdup fluctuations. Entrance effect and wall influence are also discussed in this section. Section 4.3 investigates liquid holdup distributions in the packed column. A new liquid holdup correlation is developed using the local CFD data to reduce the cost.

### 4.1 Mass transfer area

Using the VOF method, the volume faction $\alpha$ can be obtained in each computing cell. The gas–liquid interface is captured by creating an isosurface that corresponds to $\alpha = 0.5$. Similar to the specific area $a_p$, the gas–liquid interface area per unit volume $a_i$ is computed as:

$$a_i = A_i / V_s, \tag{8}$$

where $A_i$ is the gas–liquid interface area and $V_s$ is the corresponding volume. The sketch in Figure 5 (a) shows the annotations used for various phases and interfaces that can be computed and acquired in the CFD simulations of the random packed column. The white color represents the solid region, which is the pall ring in this study. The grey color stands for the gas phase, and the dark blue stands for the liquid phase in post-processed simulation results. The yellow line represents an example of a gas–liquid interface, which can be formed by the entrained droplet in gas, free streams, and liquid films on the pall ring surface. The red line represents an example of the wetted surface area, which is the contact region between the liquid and the pall ring surface. In the following discussion, we will call $a_i$ as the interfacial area by omitting per unit volume for simplicity. An example of CFD-calculated results is shown in Figure 5 (b), which shows a small portion of the packed column. The zoomed view depicts the gas–liquid flow pattern with liquid load of $q_L = 73.3$ m$^3$/m$^2$/h at $T = 5$s. From this figure, one can find how the liquid–gas interface is created in the countercurrent flow. The yellow line highlighted in the image is an example of a gas–liquid interface created by the free liquid stream, and the red line shows an example of a ring cross section area that is fully wetted by the solvent.

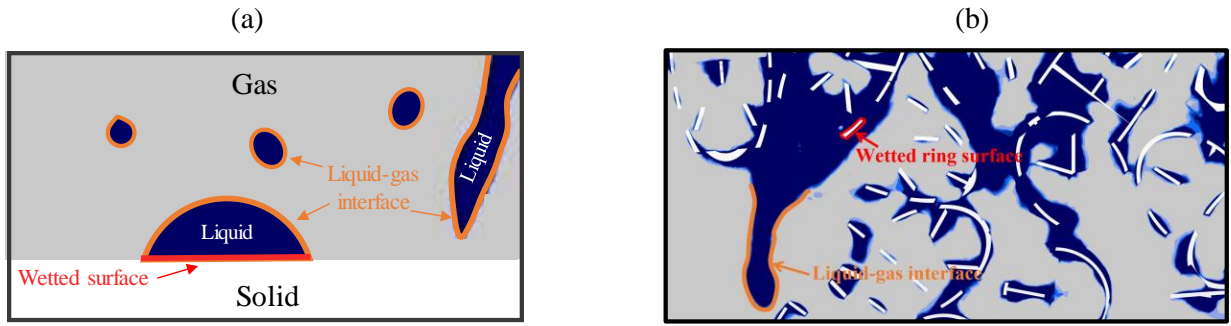

Figure 5: (a) A sketch demonstrating different regions and interfaces in CFD post-processed images. The blue and grey color denote the liquid and gas phase respectively. The gas–liquid interface (interfacial area) and liquid–solid interface (wetted area) are marked with yellow and red, respectively; (b) Exploded view of the column center section shows the countercurrent flow pattern in the random packed column with $q_L = 73.3$ m$^3$/m$^2$/h at $T = 5$s.

In the experiment, it is challenging to measure $a_i$ directly. Instead of measuring $a_i$, the effective mass transfer area $a_e$ can be backed out in an experiment based on $CO_2$ absorption rate and the measured mass transfer coefficient $k'_g$ as follows:

$$a_e = \frac{u_G}{k'_g ZRT} \ln\left(\frac{C_{CO_2,in}}{C_{CO_2,out}}\right), \tag{9}$$

where $Z$ is the height of the packed column, $R$ is the gas constant, and $T$ is temperature. $C_{CO_2,in}$ and $C_{CO_2,out}$ are the averaged $CO_2$ concentrations at the inlet and outlet of the packed column, respectively. More details of calculating $a_e$ can be found in the literature [17],[18],[31],[32]. Since the reaction generally occurs at the liquid–gas interface, it can be assumed that the liquid–gas interfacial area should be comparable to the effective mass transfer area calculated by Eq. (9). The CFD-computed $a_i$ is then compared with the correlation proposed by Song et al. [17] for validation purposes. The effective transfer area $a_e$ correlation is given as:

$$\frac{a_e}{a_p} = 1.16\eta\left[\left(\frac{\rho_L}{\sigma}\right)g^{1/2}u_L a_p^{-3/2}\right]^{0.138}, \tag{10}$$

where $a_p$ is specific area, $\sigma$ is liquid surface tension, $u_L$ is superficial liquid velocity, and $\eta$ is the coefficient, which depends on the packing design and packing material of the column. For a stainless-steel random ring packed column, $\eta$ can be calculated by:

$$\eta = 1.34 - 0.26\left(\frac{a_P}{250}\right) \tag{11}$$

To eliminate the entrance and outlet boundary influence on interfacial area $a_i$, the CFD results are calculated using the results in the height range of $z$ = [8 16] cm. The comparison between the CFD-computed $a_i$ and the correlation is shown in Figure 6. Note that the area in Figure 6 is all normalized by the total column-specific area $a_p$. Overall, the CFD liquid–gas interfacial area is comparable to the effective mass transfer area predicted by Eq. (10). The interfacial area increases with the increase of the liquid load $q_L$. In terms of magnitude, the CFD-predicted interfacial area $a_i$ is slightly underestimated compared to $a_e$ given by the correlation.

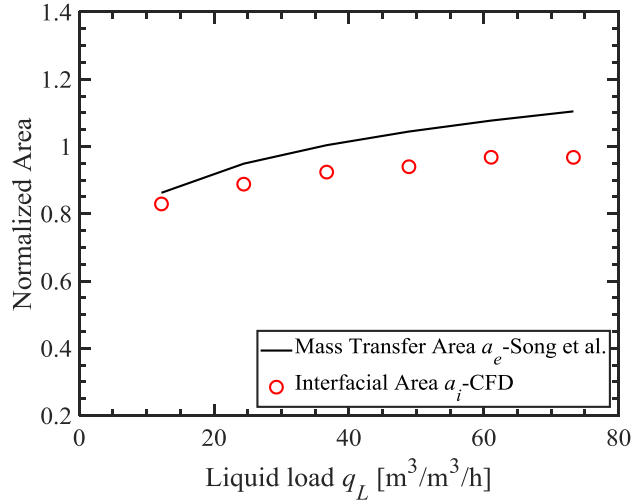

Figure 6: Comparison of the CFD-predicted interfacial area with the empirical mass transfer area given by Song et al. [17].

### 4.2 The effect of pall ring distribution

To understand the complex hydrodynamics in the packed column, the local porosity $\epsilon$ and specific area $a_p$ of the packed column are closely related to the pall ring distribution in the packed column and should be well understood. The radial porosity $\epsilon$ distribution is first plotted in Figure 7(a) against the dimensionless radius $r/R$. At the near-wall region, the $\epsilon$ has the maximum value near 1. Since the column wall surface cannot be penetrated, the pall rings can only have line or point contact with the wall. This contact mode results in a large porosity value at $r/R = 1$. Moving towards the column center with smaller $r/R$, the porosity shows a periodic fluctuation with magnitude dampened at near center region ($r/R = 0$). Two local maximum peaks are observed at $r/R = 0.08$ and 0.52. Two local minimums are located at around $r/R = 0.25$ and 0.7. The overall all packed column porosity is $\epsilon = 0.97$ in the current setup.

The specific area $a_p$ radial distribution is shown in Figure 7(b) against the dimensionless radius $r/R$. The $a_p$ distribution is also affected by the column wall and has an opposite trend compared to the porosity $\epsilon$. At near wall region ($r/R = 1$), the $a_p$ has the minimum value of 40 $m^2/m^3$. At $r/R = 0.25$ and 0.7, the $a_p$ reaches its local maximum, which is opposite to the porosity profile. The smaller porosity indicates the local region is occupied by more pall rings, which provide more surface area and increases the $a_p$ accordingly.

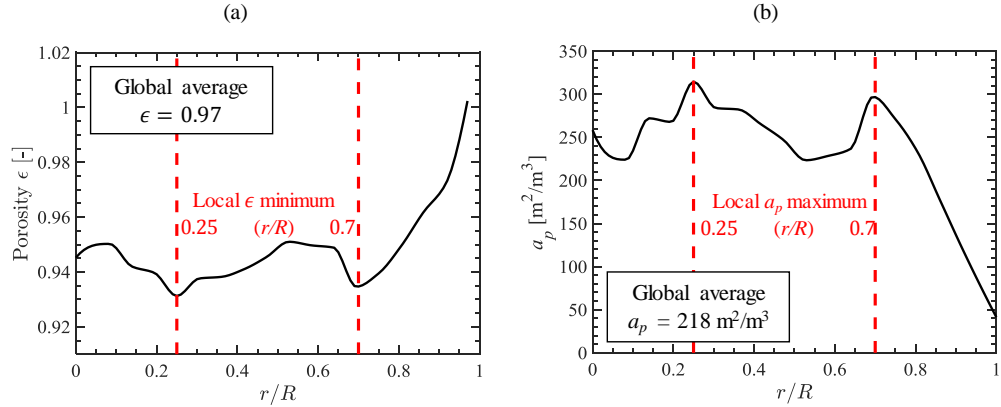

Figure 7: Geometrical properties of the random packed column: (a) radial distribution of the porosity $\epsilon$, (b) radial distribution of the specific area $a_p$.

The gas is injected at the column bottom and flows upward vertically. The gas velocity magnitude distribution at the center cross section plane is visualized in Figure 8 with liquid load $q_L$ =12.2, 48.9, and 73.3 m³/m²/h, respectively. As shown in the images, the gas velocity is bounded at 0.4 m/s in most regions. A high gas velocity can be observed in the regions with the absence of pall rings. The maximum velocity reaches 2 m/s, and these regions are mostly located in the near-wall region for all three cases. The radial superficial gas velocity $u_G$ profiles for the six runs are shown in Figure 9. Overall, the six profiles share similar trends and magnitude regardless of the liquid load. By referring to the porosity profile in Figure 7 (a), one can better see the ring distribution effect on the gas velocity $u_G$ fluctuations. As discussed in Figure 7, the near-wall region has larger porosity. With the increase of porosity, the ring number and surface decreases. This results in less friction force on the gas phase, which accordingly increases the gas velocity. This also explains the existence of larger velocity regions in Figure 8. The only discrepancy between the porosity profile and the $u_G$ is in the near-container wall region. Along with the ring surface, the column wall can create additional friction force on the gas phase. This results in a sudden gas velocity drop at the near-wall region ($r/R = 1$).

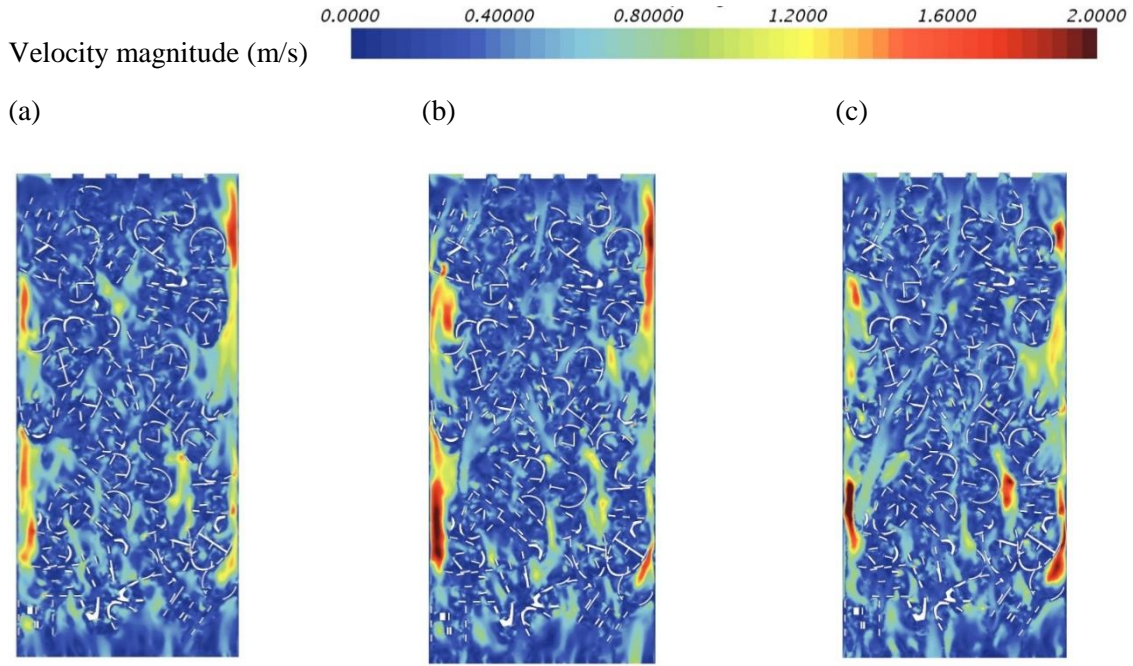

Figure 8: Gas velocity magnitude on the center cross section of the packed column at liquid flow rates of 12.2, 48.9, and 73.3 m$^3$/m$^2$/h.

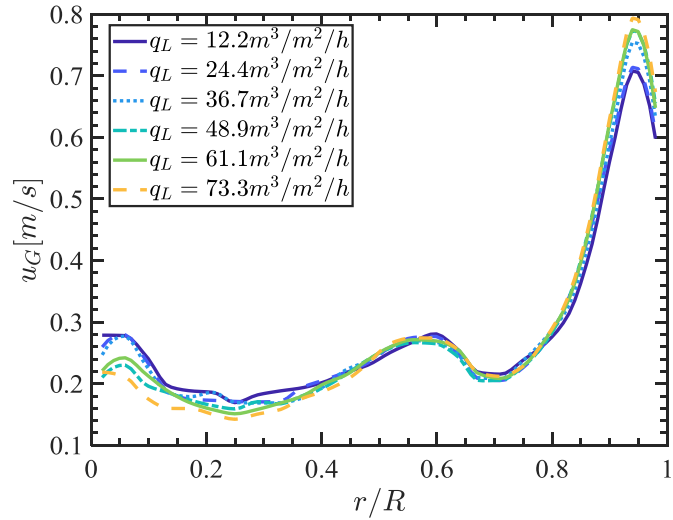

Figure 9: Radial superficial gas velocity $u_G$ distribution at various liquid load conditions.

The radial superficial liquid velocity $u_L$ distribution is plotted in Figure 10. Unlike the superficial gas velocity, the radial superficial liquid velocity $u_L$ has a clear dependence on the inlet liquid load. The large liquid load will generally result in a larger local $u_L$ across the whole column. Looking at the $a_p$ in Figure 7 (b), one can see that $u_L$ shares a similar trend to $a_p$. With a larger ring surface area, more solvent can be held at the local region, and this increases the $u_L$ accordingly. Some fluctuations are observed at the column center region for high liquid load cases ($q_L \geq 48.9$ m$^3$/m$^2$h). With a high liquid load, the solvent may not be able to disperse uniformly quick enough, which causes a local $u_L$ increase at the center region ($r/R = 0$). Besides the ring surface, the column wall can also hold a portion of the solvent. This results in an increase of $u_L$ at the near-wall region; even the $a_p$ has a decreased trend at $r/R = 1$.

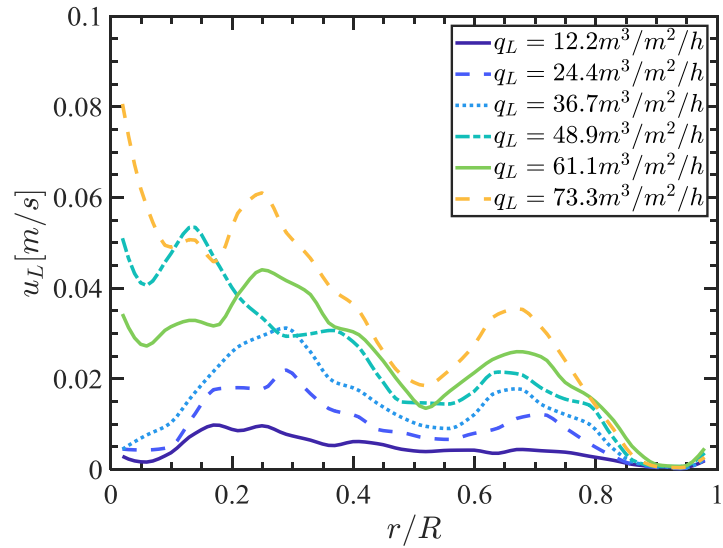

Figure 10: Radial superficial liquid velocity $u_L$ distribution at various liquid load conditions.

## 4.3 Development of liquid holdup correlation

The liquid holdup $h_L$ can significantly influence the hydrodynamics of a packed column and is used to estimate the requested amount of solvent used for carbon capture. The wet pressure drop calculation across the column also relies on the accurate prediction of the liquid holdup in the column [33]. Here, we extensively investigated the local and global liquid holdup in the packed column. The local liquid holdup distributions are analyzed at different heights ($z$) and radial positions ($r/R$). The solvent maldistribution and entrance effect are observed in the random packed column. A new correlation is proposed for liquid holdup prediction after analyzing its dependency on the specific area $a_p$ and liquid velocity $u_L$ profile. The global liquid holdup results and empirical correlations are compared with the proposed correlation. By definition, the liquid holdup is calculated as the fraction of liquid volume ($V_L$) over the total empty column volume ($V_S$) as:

$$h_L = V_L / V_S. \tag{12}$$

To investigate the entrance effect, the column is divided into ten sections uniformly along vertical direction $z$. In each section, the radial liquid holdup profile is plotted for visualization and analysis. In order to generate the radial profile for each section, $h_L$ is averaged along the angular direction in polar coordinates as:

$$h_L(r) = \frac{\int_0^{2\pi} h_L(r,\theta) d\theta}{2\pi}. \tag{13}$$

The radial liquid holdup profile at different sections is plotted in Figure 11 using the $q_L$ = 36.7 m³/m²/h case. To better visualize the entrance effect influence on the liquid holdup distribution, horizontal sections are divided into two plots. Figure 11 (a) shows the top two sections $z_9$ ([16–18 cm]) and $z_{10}$ ([18–20 cm]), which are close to the liquid inlets. The rest of the sections in between $z$ = 0 cm and 16 cm are visualized in Figure 11 (b). The red vertical dashed lines in Figure 11 (a) represent the locations of the three inlet drip groups at $r/R$ = 0.3, 0.42, and 0.6. The red vertical dashed lines in Figure 11 (b) correspond to the local $a_p$ maximums found in Figure 7 (b). By examining the top section $z_{10}$, one can see that there is a strong entrance effect right below the inlets. The three local liquid holdup peaks appear exactly at the locations of the three inlet hole sets. Due to the relatively small column diameter and reasonable inlet coverage, the entrance effect disappeared shortly. The peaks in the top section $z_{10}$ have been smoothed out in section $z_9$. The center peaked is dispersed, and a peak begins to grow near $r/R = 0.7$. For the height $z$ below 16 cm, the solvent is well spread. The radial liquid holdup profiles shown in Figure 11 (b) form two new peaks at around $r/R = 0.25$ and 0.7. These locations correspond to the local $a_p$ maximum, as shown in Figure 7 (b). The packed column local morphology is shown to have a more dominant influence at those regions with $z < 16$ cm.

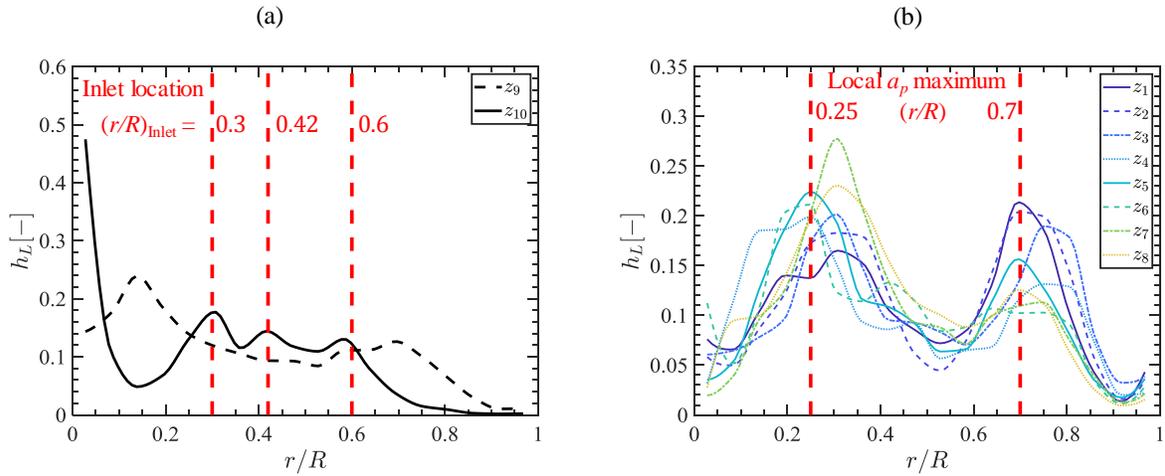

Figure 11: The radial liquid holdup distribution with the liquid load at $q_L = 36.7$ m$^3$/m$^2$/h. The whole packed column is divided into ten sections: (a) sections $z_9$ and $z_{10}$, which are close to the liquid inlets, and (b) sections $z_1$–$z_8$, which cover heights between 0 cm and 16 cm.

To further examine the impact of the liquid flow rate on solvent distribution, the radial $h_L$ profiles for six different cases are plotted in Figure 11. Liquid holdup $h_L$ is averaged along the $z$ direction for the whole column. The top two sections ($z_1$, $z_2$) are excluded to eliminate the entrance effect influence. Comparing the six cases, one can see that the liquid holdup increases with the increase of liquid load. The two local $h_L$ peaks are more concentrated at around $r/R = 0.25$ and 0.7, which correspond to the local $a_p$ peaks. Since the solvent formed the liquid films on the ring surface, the larger local surface area will hold more solvent in that region. This results in an increase in local liquid holdup $h_L$ with a constant inlet liquid load.

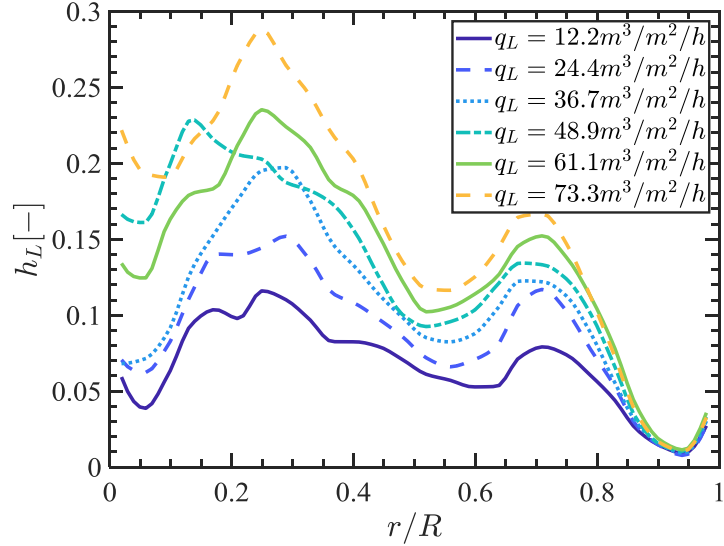

Figure 12: Local radial liquid holdup $h_L$ distribution for six liquid load cases. Liquid holdup $h_L$ is averaged along the $z$ direction.

The liquid holdup data from CFD simulations are further investigated for developing a correlation to predict the liquid holdup. In our current CFD simulations, only six runs are carried out. The limited column-averaged liquid holdup data are not sufficient for the fitting correlation, which can have a strong dependence on both liquid velocity $u_L$ and specific area $a_p$. To address this issue, an alternative method is proposed to develop the liquid holdup correlation in the pall ring packed column by using local geometrical and hydrodynamics data acquired in the simulations.

The existing correlations are first summarized and briefly discussed in Table 2. In brief, the percentage of liquid holdup is determined by three forces, namely, gravity $F_g$, viscous force $F_\mu$, and resistance force $F_\psi$ [33]. Gravity is the driving force and is balanced by the combination of viscous force $F_\mu$ and resistance force $F_\psi$. Depending on the Re, the weight of $F_\mu$ and $F_\psi$ can be different. With a small Re, the viscous force is dominant and the liquid holdup follows a power law of $h_L \sim u_L^{1/3}$, as described by

Mackowiak [34] and Bemer and Kalis [33]. For a turbulent regime and large Re value, the resistance force is dominant over the viscous force. Using force balance, Bemer and Kalis [33] derived a theoretical model to indicate that the liquid holdup follows a power law $h_L \sim u_L^{2/3}$ with an exponent of 2/3.

Table 2: Correlations for liquid holdup in a packed column.

| Model | Correlation for $h_L$ | Remarks |
|---|---|---|
| **Buchanan (1969)[35]**[1] | $h_L = 2.2\left(\dfrac{Fr'}{Re'}\right)^{1/3} + 1.8 Fr^{1/2} = 2.2\left(\dfrac{u_L \mu_L}{gd^2 \rho_L}\right)^{1/3} + 1.8\left(\dfrac{u_L^2}{gd}\right)^{1/2}$ | Empirical correlation fitted with comprehensive experimental data |
| **Bemer and Kalis (1978)[33]** | $h_L = 0.34 a_p^{1/3} u_L^{2/3}$ | Derived by assuming resistance force dominant |
| **Suess and Spiegel (1992)[36]**[2] | $h_L = \begin{cases} 0.0169 a_p^{0.83} q_L^{0.37}\left(\dfrac{\mu_L}{\mu_W}\right)^{0.25}, q_L < 40 \text{ m}^3/\text{m}^2\text{h} \\ 0.0075 a_p^{0.83} q_L^{0.59}\left(\dfrac{\mu_L}{\mu_W}\right)^{0.25}, q_L > 40 \text{ m}^3/\text{m}^2\text{h} \end{cases}$ | Empirical correlation fitted with experimental data from air-water system |
| **Mackowiak (2010)[34]** | $h_L = \dfrac{3}{4}\left(\dfrac{3}{g}\right)^{1/3} a_p^{2/3}\left(\dfrac{\mu_L}{\rho_L}\right)^{1/3} u_L^{1/3}$ | Derived by assuming viscous force |

[1] $Fr' = \dfrac{u_L^2}{gd}$, $Re' = \dfrac{u_L d \rho_L}{\mu_L}$, $d$ = ring diameter or height. [2] $\mu_W$ refers to water viscosity.

Along with the liquid velocity $u_L$, the specific area $a_p$ and the liquid viscosity $\mu_L$ are other factors that can affect the liquid holdup, as shown in Table 2. In terms of dimensionless number, the Reynolds number $Re = \dfrac{\rho_L u_L}{a_p \mu_L}$ and Froude number $Fr = \dfrac{u_L^2 a_p}{g}$ include these aforementioned factors in liquid holdup prediction. Therefore, the liquid holdup correlation is assumed to have the form of:

$$h_L = C_1 Re^{C_2} Fr^{C_3}, \tag{14}$$

where $C_1$, $C_2$, and $C_3$ are the coefficients to be determined. Liquid density and viscosity are treated as constant in this study. Expanding Eq. (14), one can express $h_L$ in terms of $u_L$ and $a_p$ as

$$h_L = \left( \frac{C_1 \rho_L^{C_2}}{\mu_L^{C_2} g^{C_3}} \right) u_L^{C_2+2C_3} a_p^{-C_2+C_3}. \tag{15}$$

With sufficient input parameter $a_p$ and $u_L$ data, the coefficients in the correlation can be extracted accordingly. The entire column has a fixed specific area $a_p = 218$ m$^2$/m$^3$. With only six CFD runs at different liquid loads, it is not practical to regress all three coefficients in Eq. (15) accurately. However, by dividing the column into concentric cylinders and looking at the local radial CFD data, the available range of specific area $a_p$ and liquid velocity $u_L$ can be significantly expanded. A comparison of using the column-averaged and local CFD data is shown in Figure 13. As shown in the figure, with a fixed $a_p$ and six-liquid flow rate, the ($u_L$, $a_p$) input parameter space is very narrow using the column-averaged data. The covered liquid holdup range is [0.05–0.12]. With local CFD data, the radial $a_p$ can cover a range of [40–315] m$^2$/m$^3$, and the radial liquid velocity $u_L$ can cover a range of [0–0.08] m/s. The corresponding liquid holdup $h_L$ profile has a range of [0–0.3], which is four times larger than the column-averaged liquid holdup range. Detailed radial $a_p$, $u_L$, and $h_L$ profiles can be referred to in Figure 7 (b), Figure 10, and Figure 12, respectively.

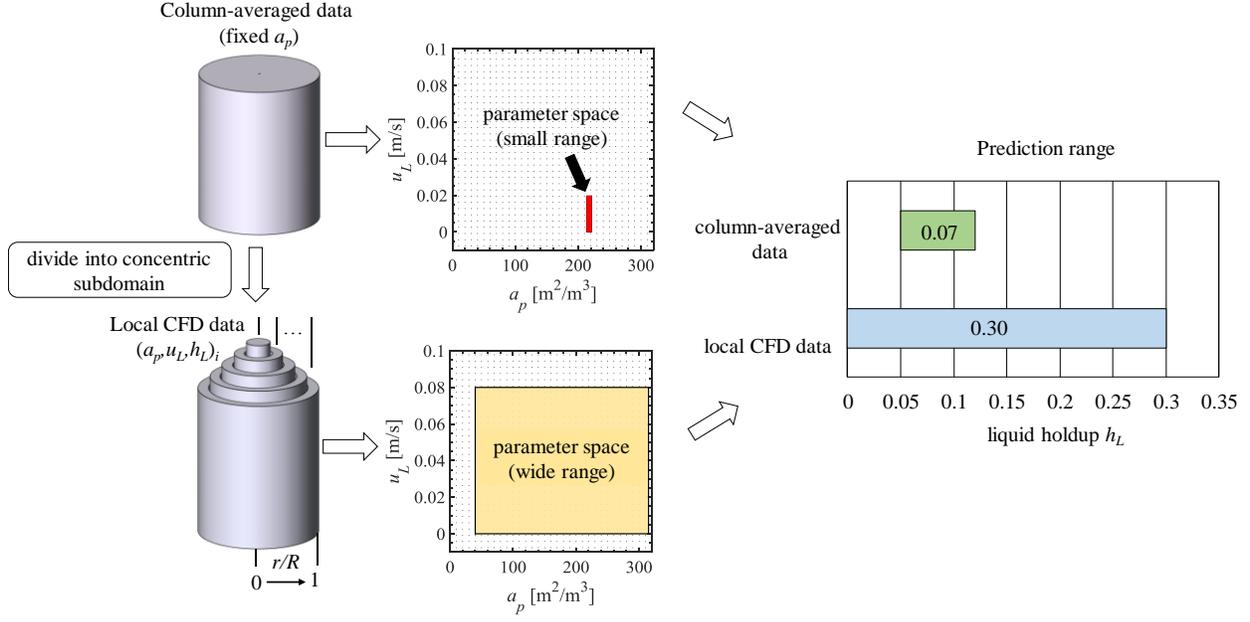

Figure 13: Comparison of input parameter (specific area $a_p$, liquid velocity $u_L$) range using the column-averaged and local CFD data method. The corresponding liquid holdup range using the two methods is demonstrated in the bar plot.

With the expanded $u_L$, $a_p$, and $h_L$ data range in local CFD data, the exponent of $u_L$ and $a_p$ in Eq. (15) can be fitted using the multiple linear regression method [37]. In order to perform the multiple linear regression, Eq. (15) is reformed as

$$\ln(h_L) = \ln\left(\frac{C_1 \rho_L^{C_2}}{\mu_L^{C_2} g^{C_3}}\right) + (C_2 + 2C_3)\ln u_L + (-C_2 + C_3)\ln a_p. \tag{16}$$

For each CFD run, 14 data points are sampled uniformly along the radial direction for $a_p$, $u_L$, and $h_L$, respectively. The data near the $r/R = 1$ region are excluded to remove the wall influence. With six CFD runs, a total of 84 ($a_p$, $u_L$, $h_L$) pairs were extracted from the simulations. When these data are plugged into Eq. (16), the three coefficients are calculated as:

$$C_1 = 7.81, C_2 = -0.52, C_3 = 0.48. \tag{17}$$

Using these coefficients, the liquid holdup correlation can be derived as:

$$h_L = 7.81 \mathrm{Re}^{-0.52} Fr^{0.48} = 7.81 \frac{1}{g^{0.48}} \left(\frac{\mu_L}{\rho_L}\right)^{0.52} a_p u_L^{0.44}. \tag{18}$$

The comparison of the developed correlation with local CFD liquid holdup data is shown in Figure 14. The developed correlation is validated against the liquid holdup range of [0–0.3]. The correlation can predict the local liquid holdup results with an error percentage of ±15%.

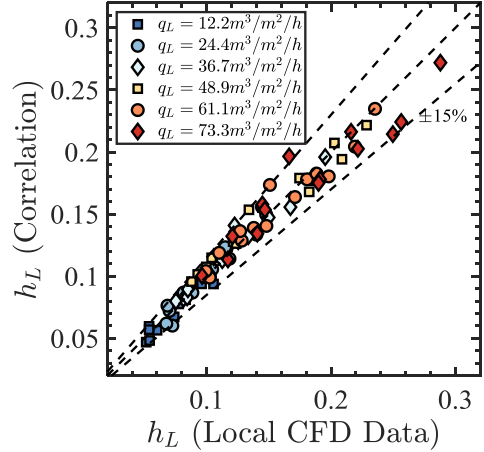

Figure 14: Comparison of local CFD liquid holdup data with the proposed correlation at six liquid load conditions.

Since the acquired correlation is derived based on the local CFD data, the applicability of this correlation for predicting the column-averaged liquid holdup results should be further validated. This was achieved by comparing Eq. (18) with the column-averaged liquid holdup data along with four existing correlations. The results are shown in Figure 15. As shown in the figure, the proposed correlation using the local CFD data has a good agreement with the global column-averaged $h_L$ at various liquid load conditions. The exponent of $u_L$ in the correlations is annotated in the plot. In the current study, the exponent of $u_L$ is regressed as 0.44. The exponent value 0.44 is close to the correlations of $h_L \sim u_L^{5/11}$ given by Gelbe [38] and of $h_L \sim u_L^{0.45}$ given by Mersmann and Deixler [39]. This value falls in between the viscous regime with exponent of $h_L \sim u_L^{1/3}$ [34] and the turbulent regime with exponent of $h_L \sim u_L^{2/3}$ [33]. For Suess and Spiegel's correlation, the equation indicates that transition happens at $q_L = 40$ m$^3$/m$^2$/h. With a small liquid load, the liquid holdup follows the power law of $h_L \sim u_L^{0.37}$. With a large liquid load, the liquid holdup follows the power law of $h_L \sim u_L^{0.59}$. This evidence further confirms that the current countercurrent flow in the packed column is a transition regime with a Reynold number in between 6.7 and 40.2. The viscous force and resistance force have similar orders of magnitude and cannot be omitted at this transition regime.

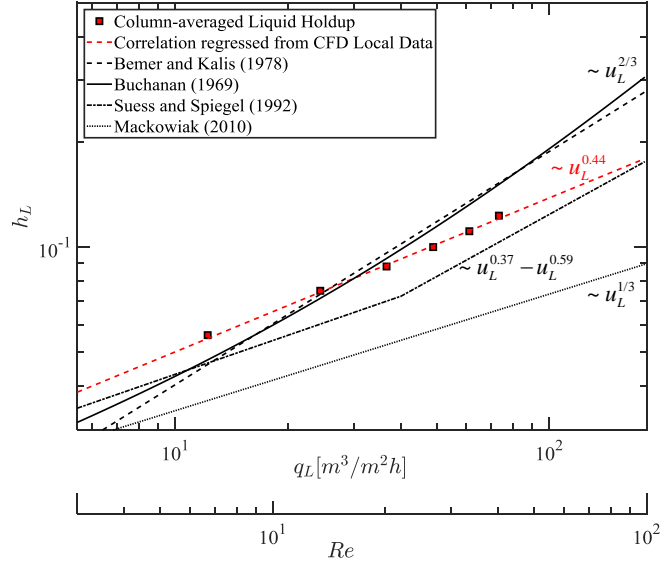

Figure 15: Comparison of proposed liquid holdup correlation (based on CFD local data) with column-averaged $h_L$ and with existing liquid holdup correlations.

## 5 Conclusion

A high-fidelity CFD approach is developed to investigate countercurrent multiphase flows in a random packed column. The proposed approach can predict the liquid holdup and mass transfer area for packed column performance assessment. Both the column-scale and local characteristics of the countercurrent flow are visualized and investigated. The radial profiles of $h_L$, $u_L$, and $u_G$ are found to be closely related to the local packed column porosity $\epsilon$ and specific area $a_p$. The wall influence on the packed column velocity and liquid holdup are observed in this study. Additional wall surface area can hold more liquid and thus increases the near-wall liquid velocity $u_L$. However, the wall can also increase the friction force, which decreases the gas velocity $u_G$. The entrance effect is observed at the near-inlet locations, but this effect disappeared after around 4 cm using the 13-inlet drip hole setup at a moderate liquid load. After 4 cm of development, the liquid holdup distribution will be reshaped to closely follow the specific area $a_p$ profile.

In terms of the gas–liquid interface, the proposed CFD model has the capability of visualizing and calculating the interfacial area by extracting the isosurface with void fraction $\alpha = 0.5$. The liquid–gas interface area is comparable to the existing empirical correlation. It should be noticed that the interfacial area is derived directly using the void fraction information and does not require additional assumptions. This indicates that the proposed procedure can be extended to different types of random packed columns for gas–liquid interfacial area prediction.

A new liquid holdup correlation is proposed by using 84 pairs of local ($a_p$, $u_L$, $h_L$) data. In the traditional method, a large amount of CFD simulations or experiments should be carried out using various packing types and different liquid loads in order to understand the impact of $a_p$ and $u_L$ on $h_L$. By using the local CFD data in the packed column, the liquid holdup correlation can be regressed successfully using only one size of pall ring and limited simulation runs. The developed correlation can predict local liquid holdup within $\pm 15\%$ error. The applicability of this methodology is further validated by comparing it with the column-averaged liquid holdup results and existing correlations. The relationship between $u_L$ and $h_L$ is discussed in detail. The proposed correlation shows that $h_L$ follows a power law of $h_L \sim u_L^{0.44}$ in the Reynolds number range of Re =[6.7–40.2]. This shows that for 16 mm pall rings, the packed column falls into the transition regime where both viscous force and inertial force have equal importance.

**Acknowledgments**

Pacific Northwest National Laboratory is operated by Battelle for the U.S. Department of Energy (DOE) under Contract No. DE-AC05-76RL01830. This work was funded by the DOE Office of Fossil Energy's Carbon Capture Simulation Initiative (CCSI) through the National Energy Technology Laboratory.

**References**

[1]  B. Metz, O. Davidson, H. de Coninck, M. Loos, and L. Meyer, "IPCC Special Report on Carbon Dioxide Capture and Storage," 2005.

[2]  R. S. Haszeldine, "Carbon Capture and Storage: How Green Can Black Be?," *Science*, vol. 325, no. 5948, pp. 1647–1652, 2009.

[3]  K. Sumida *et al.*, "Carbon dioxide capture in metal–organic frameworks," *Chem. Rev.*, vol. 112, no. 2, pp. 724–781, 2011.

[4]  Z. Chen, D. Yates, J. K. Neathery, and K. Liu, "The effect of fly ash on fluid dynamics of $CO_2$ scrubber in coal-fired power plant," *Chem. Eng. Res. Des.*, vol. 90, no. 3, pp. 328–335, 2012.

[5]  P. L. Spedding, M. T. Jones, and G. R. Lightsey, "Ammonia absorption into water in a packed tower II: Measurement of mass transfer coefficients," *Chem. Eng. J.*, vol. 32, no. 3, pp. 151–163, 1986.

[6]  R. J. Kouri and J. Sohlo, "Liquid and gas flow patterns in random packings," *Chem. Eng. J. Biochem. Eng. J.*, vol. 61, no. 2, pp. 95–105, 1996.

[7]  F. Heymes, P. Manno Demoustier, F. Charbit, J. Louis Fanlo, and P. Moulin, "Hydrodynamics and mass transfer in a packed column: Case of toluene absorption with a viscous absorbent," *Chem. Eng. Sci.*, vol. 61, no. 15, pp. 5094–5106, 2006.

[8]  P. Alix and L. Raynal, "Liquid Distribution and Liquid Hold-up in a high capacity structured packing," *Chem. Eng. Res. Des.*, vol. 86, pp. 1–6, 2008.

[9]  S. Piché, F. Larachi, and B. P. A. Grandjean, "Improved liquid hold-up correlation for randomly packed towers," *Chem. Eng. Res. Des.*, vol. 79, no. 1, pp. 71–80, 2001.

[10]  S. R. Piché, F. Larachi, and B. P. A. Grandjean, "Improving the prediction of irrigated pressure drop in packed absorption towers," *Can. J. Chem. Eng.*, vol. 79, no. 4, pp. 584–594, 2001.

[11]  S. Piché, F. Larachi, and B. P. A. Grandjean, "Loading capacity in packed towers - Database, correlations and analysis," *Chem. Eng. Technol.*, vol. 24, no. 4, pp. 373–380, 2001.

[12]  K. Onda, H. Takeuchi, and Y. Okumoto, "Mass transfer coefficients between gas and liquid

phases in packed columns," *J. Chem. Eng. Japan*, vol. 1, no. 1, pp. 56–62, 1968.

[13] Z. Xu, R. K. Singh, J. Bao, and C. Wang, "Direct Effect of Solvent Viscosity on the Physical Mass Transfer for Wavy Film Flow in a Packed Column," *Ind. Eng. Chem. Res.*, vol. 58, pp. 17524–17539, 2019.

[14] Y. Son and W. Won, "Liquid holdup and pressure drop in packed column with structured packing under offshore conditions," *Chem. Eng. Sci.*, vol. 195, pp. 894–903, 2019.

[15] R. Sidi-Boumedine and L. Raynal, "Influence of the viscosity on the liquid hold-up in trickle-bed reactors with structured packings," *Catal. Today*, vol. 105, no. 3–4, pp. 673–679, 2005.

[16] P. Marchot, D. Toye, A. M. Pelsser, M. Crine, G. L'Homme, and Z. Olujic, "Liquid distribution images on structured packing by X-ray computed tomography," *AIChE J.*, vol. 47, no. 6, pp. 1471–1476, 2001.

[17] D. Song, A. F. Seibert, and G. T. Rochelle, "Mass Transfer Parameters for Packings: Effect of Viscosity," *Ind. Eng. Chem. Res.*, vol. 57, no. 2, pp. 718–729, 2018.

[18] R. Tsai, "Mass Transfer Area of Structured Packing," The University of Texas at Austin, 2010.

[19] C. Wang, "Mass Transfer Coefficients and Effective Area of Packing," The University of Texas at Austin, 2015.

[20] J. L. Kang, W. F. Chen, D. S. H. Wong, and S. S. Jang, "Evaluation of gas-liquid contact area and liquid holdup of random packing using CFD simulation," *2017 6th Int. Symp. Adv. Control Ind. Process. AdCONIP 2017*, pp. 636–641, 2017.

[21] T. Eppinger, K. Seidler, and M. Kraume, "DEM-CFD simulations of fixed bed reactors with small tube to particle diameter ratios," *Chem. Eng. J.*, vol. 166, no. 1, pp. 324–331, 2011.

[22] Y. Dong, B. Sosna, O. Korup, F. Rosowski, and R. Horn, "Investigation of radial heat transfer in a fixed-bed reactor: CFD simulations and profile measurements," *Chem. Eng. J.*, vol. 317, pp. 204–214, 2017.

[23] S. Rebughini, A. Cuoci, and M. Maestri, "Handling contact points in reactive CFD simulations of heterogeneous catalytic fixed bed reactors," *Chem. Eng. Sci.*, vol. 141, pp. 240–249, 2016.

[24] E. Zaman and P. Jalali, "On hydraulic permeability of random packs of monodisperse spheres: Direct flow simulations versus correlations," *Phys. A Stat. Mech. its Appl.*, vol. 389, no. 2, pp. 205–214, 2010.

[25] H. P. A. Calis, J. Nijenhuis, B. C. Paikert, F. M. Dautzenberg, and C. M. Van Den Bleek, "CFD modeling and experimental validation of pressure drop and flow profile in a novel structured catalytic reactor packing," *Chem. Eng. Sci.*, vol. 56, no. 4, pp. 1713–1720, 2001.

[26] D. Sebastia-Saez, S. Gu, and M. Ramaioli, "Effect of the contact angle on the morphology, residence time distribution and mass transfer into liquid rivulets: A CFD study," *Chem. Eng. Sci.*, vol. 176, pp. 356–366, 2018.

[27] N. Jurtz, M. Kraume, and G. D. Wehinger, "Advances in fixed-bed reactor modeling using particle-resolved computational fluid dynamics (CFD)," *Rev. Chem. Eng.*, vol. 35, no. 2, pp. 139–190, 2019.

[28] R. Caulkin, A. Ahmad, M. Fairweather, X. Jia, and R. A. Williams, "Digital predictions of complex cylinder packed columns," *Comput. Chem. Eng.*, vol. 33, no. 1, pp. 10–21, 2009.

[29] R. Caulkin, X. Jia, M. Fairweather, and R. A. Williams, "Lattice approaches to packed column simulations," *Particuology*, vol. 6, no. 6, pp. 404–411, 2008.

[30] "CD-adapco STAR-CCM+ 13.04 User Guide." 2018.

[31] C. Wang, Z. Xu, C. Lai, and X. Sun, "Beyond the standard two-film theory: Computational fluid dynamics simulations for carbon dioxide capture in a wetted wall column," *Chem. Eng. Sci.*, vol. 184, pp. 103–110, 2018.

[32] C. Wang, Z. Xu, K. Lai, G. Whyatt, P. W. Marcy, and X. Sun, "Hierarchical calibration and validation framework of bench-scale computational fluid dynamics simulations for solvent-based carbon capture. Part 2: Chemical absorption across a wetted wall column," *Greenh. Gases Sci. Technol.*, vol. 8, no. 1, pp. 150–160, 2018.

[33] G. G. Bemer and G. A. J. Kalis, "A new method to predict Hold-up and pressure drop in packed columns," *Trans IChemE*, vol. 56, pp. 200–204, 1978.

[34]     J. Mackowiak, *Fluid Dynamics of Packed Columns*. Springer Berlin Heidelberg, 2010.

[35]     J. E. Buchanan, "Pressure gradient and liquid holdup in irrigated packed towers," *Ind. Eng. Chem. Fundam.*, vol. 8, no. 3, pp. 502–511, 1969.

[36]     P. Suess and L. Spiegel, "Hold-up of mellapak structured packings," *Chem. Eng. Process.*, vol. 31, no. 2, pp. 119–124, 1992.

[37]     R. K. Singh, J. E. Galvin, and X. Sun, "Multiphase flow studies for microscale hydrodynamics in the structured packed column," *Chem. Eng. J.*, vol. 353, no. July, pp. 949–963, 2018.

[38]     H. Gelbe, "Rektifizierwirkung bei Vakuumbetrieb in Füllkörperschüttungen," *Chemie Ing. Tech.*, vol. 40, no. 11, pp. 528–530, Jun. 1968.

[39]     A. Mersmann and Alfred Deixler, "Packungskolonnen," *Chemie Ing. Tech.*, vol. 58, no. 1, pp. 19–31, 1986.

***Declaration of Interest Statement**

**Declaration of interests**

☒ The authors declare that they have no known competing financial interests or personal relationships that could have appeared to influence the work reported in this paper.

☐ The authors declare the following financial interests/personal relationships which may be considered as potential competing interests:

Yucheng Fu,
Jie Bao,
Rajesh Singh,
Chao Wang,
Zhijie Xu